\documentclass{caosp}

\usepackage{graphicx}

\articleNo{123}
\pubyear{2005}
\volume{35}
\volnumber{3}
\firstpage{1}
\received{May 1, 2005}
\accepted{August 28, 2005}

\begin{document}

\hauthor{O. Kochukhov and D. Shulyak}

\title{New generation model atmospheres for chemically peculiar stars}


\author{O. Kochukhov \inst{1} \and D. Shulyak \inst{2}}

\institute{Department of Astronomy and Space Physics, 
           Uppsala University, Box 515, SE-751 20 Uppsala, Sweden
         \and 
         Institute for Astronomy, University of Vienna, 
         T\"urkenschanzstra{\ss}e 17, A-1180 Vienna, Austria
          }

\date{December 1, 2007}

\maketitle

\begin{abstract}
The atmospheric structure of chemically peculiar stars deviates from that of normal
stars with similar fundamental parameters due to unusual chemistry, abundance
inhomogeneities and the presence of strong magnetic field. These effects are not considered
in the standard model atmospheres, possibly leading to large errors in
the stellar parameter determination and abundance analysis. To tackle this problem
we used the state-of-the-art opacity sampling model atmosphere code {\tt LLmodels} to
calculate comprehensive grid of new generation model atmospheres for magnetic CP
stars. This grid 
covers the whole parameter space occupied by SrCrEu and Si-peculiar stars, taking
into account characteristic temperature dependence of the chemical abundances.
Here we present the first results of our model atmosphere calculations. 
\keywords{stars: atmospheres -- stars: chemically peculiar}
\end{abstract}

\section{Model atmosphere calculations}

To calculate realistic models of chemically peculiar stars we have used 
a 1-D, LTE, hydrostatic model atmosphere code {\tt LLmodels} (Shulyak. et al. \cite{llmodels}). This
code allows one to treat the line opacity without simplifying approximations and in
this way accurately estimate how modification of the line opacity due to anomalous abundances,
chemical stratification or the presence of strong magnetic field affects stellar
atmospheres. 

Our calculations cover the $T_{\rm eff}$ range between 6500 and 18000~K
with a step of 250--500~K and the $\log g$ range of 3.25--4.5~dex with a step of
0.25~dex. For $\log g=4.0$ models we have performed additional calculations for 
5~kG magnetic field using the method described by Kochukhov et al. (\cite{llmag}).
Individual abundances of 41 elements were adopted based on the results of modern
chemical composition analyses of the SrCrEu and Si-type CP stars. For
Ca, Si, Cr, Fe and several other elements we include empirical temperature dependence
of the element concentrations (Ryabchikova \cite{r05}). The line lists are extracted
from the VALD database (Kupka et al. \cite{vald}). The Ap-star model grid is
complemented with the solar composition models, calculated for an extended range
of atmospheric parameters (6000--20000~K in $T_{\rm eff}$ and 2.5--4.5~dex in $\log g$).

\section{Results}

\begin{figure}[!t]
\centerline{\includegraphics[width=12cm]{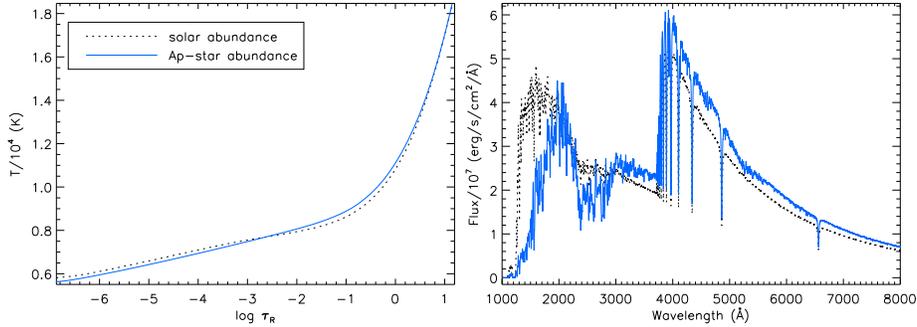}}
\caption{Results of the model atmosphere calculations for a star with 
$T_{\rm eff}=10000$~K and $\log g=4.0$. The $T$-$\tau$ relations (left panel) and 
flux distributions (right panel)
are presented for the 
models with solar abundances 
and the Ap-star abundance table.}
\label{fig1}
\end{figure}

For each Ap-star model atmosphere we have examined deviation of the temperature
stratification, flux distribution, hydrogen line profiles and photometric colors from
those of normal composition model with the same parameters. Fig.~\ref{fig1} illustrates
the impact of anomalous abundances on the atmospheric structure of
an early-A star. In general, we find that\\
(i) peculiar abundances lead to 200--400~K increase of $T$ in deep layers and 
100--200~K cooling of the upper atmosphere;\\
(ii) the impact of 5~kG magnetic field is negligible compared to that of the chemical composition
anomalies;\\
(iii) fitting normal composition models to the hydrogen lines of Ap stars
underestimates $T_{\rm eff}$ by $\approx$\,250~K and $\log g$ by $\approx$\,0.25~dex;\\
(iv) model fluxes of Ap stars are dominated by the energy redistribution from UV to
visual and IR wavelengths; the bolometric corrections of Ap and normal stars differ
systematically by $\le0.1$~mag;\\
(v) usual photometric calibrations overestimate $T_{\rm eff}$ of Ap stars by
300--500~K. 

\acknowledgements
This work was supported by the
Austrian Science Fund (FWF) grant P17890 to OK
and the FWF Lisa Meitner grant M998-N16 to DS.

\end{document}